\documentclass[usegraphicx,usenatbib,useapjfonts,apj]{emulateapj}

\shorttitle{$H_{2}$, $HD$ and $LiH$ cooling in primordial gas.}
\shortauthors{Prieto et al.}

\begin{document}


\title{Revisiting the effect of $H_{2}$, $HD$ and $LiH$ molecules\\ in the cooling of primordial gas.}


\author{Joaquin P. Prieto\altaffilmark{1}, Leopoldo Infante\altaffilmark{1}, Raul Jimenez\altaffilmark{2}}
\altaffiltext{1}{Departamento de Astronom\'ia y Astrof\'isica, Pontificia Universidad Cat\'olica de Chile, Santiago, Chile; jpprieto@astro.puc.cl, linfante@astro.puc.cl}
\altaffiltext{2}{ICREA \& Institute of Space Sciences (CSIC-IEEC), Campus UAB, Bellaterra 08193, Spain; raulj@astro.princeton.edu}



\begin{abstract}
We use a non-equilibrium chemical network to revisit and study the effect of $H_{2}$, $HD$ and $LiH$ molecular cooling on a primordial element of gas. We paid special attention in the variation of $HD$ abundance. We solve both the thermal and chemical equations for a gas element with an initial temperature $T = 1000K$ and a gas number density in the range $n_{tot}=1-10^{3} cm^{-3}$. These are typical propierties of the first halos which formed stars. At low densities, $n_{tot}<10^{2} cm^{-3}$, the gas reaches temperatures $\sim 100K$ and the main coolant is $H_{2}$, but at higher densities, $n_{tot}>10^{2} cm^{-3}$, the $HD$ molecule dominates the gas temperature evolution and the gas reaches temperatures well below $100K$. The effect of $LiH$ is negligible in all cases. We studied the effect of $HD$ abundance on the gas cooling. The $HD$ abundance was set initially to be in the range $n_{HD}/n_{H}=10^{-7}-10^{-5}$. The simulations show that at $n_{tot}>10^{2} cm^{-3}$ the HD cooling dominates the temperature evolution for $HD$ abundances greater than $10^{-6}n_{H}$. This number decrease at higher densities. Furthermore, we studied the effect of electrons and ionized particules on the gas temperature. We followed the gas temperature evolution with $n_{H_{+}}/n_{H}=10^{-4}-10^{-2}$. The gas temperature reached lower values at high ionization degree because electrons, $H^{+}$ and $D^{+}$ are catalizers in the formation paths of the $H_{2}$ and $HD$ molecules. Finaly, we studied the effect of an OB star, with $T_{eff}=4\times 10^{4}K$, would have on gas cooling. It is very difficult for a gas with $n_{tot}$ in the range between $1-100 cm^{-3}$ to drop its temperature if the star is at a distance less than $100 pc$.
\end{abstract}


\section{Introduction}
In a $\Lambda CDM$ Universe, the first luminous objetcs were formed due to fall-in of gas inside the dark matter potential wells (for a review see  \citet{Barkana 2001}). In order to let the fall-in of gas inside the dark matter potential wells, the gas thermal energy should have been radiated away by some physical mechanism. Since the primordial molecular clouds would have zero metallicity, the collisional excitation of the existing molecules is the most plausible mechanism for the cooling of baryonic matter in this environment because the collisional excitation cooling of both $H$ and $He$ atoms are inefficient at temperatures lower than $8000K$, which is higher than the temperature of star formation clouds.

\citet{Tegmark 1997} showed that the first lumious objects may have formed at $z\sim 30$ inside a $10^{6}M_{\odot}$ halo ($T_{vir}\sim 1000K$), which have recently been confirmed by \citet{O'Shea 2007} and \citet{Gao 2007}. \citet{Tegmark 1997} also showed that the stars formed in this environment could be as massives as $\sim 100M_{\odot}$ \citep{Abel2002}.

After the recombination era, the most abundant molecule in the Universe is molecular hydrogen $H_{2}$. Despite of its low primordial abundance, $\sim 10^{-3}-10^{-4}n_{H}$ \citep{Palla et al. 1983}, this molecule has a fundamental role in the gas cooling at temperatures less than $8000K$. The first authors to highlight the role of $H_{2}$ in this context were \citet{Saslaw} and \citet{Peebles 1968}. Saslaw \& Zipoy (1967) showed the importance of the charge transfer reaction between $H_{2}^{+}$ and $H$ to form $H_{2}$, and Peebles \& Dicke (1968) suggested a mechanism to form $H_{2}$ from $H^{-}$.

The $H_{2}$ molecule forms by the $H^{-}$ and $H_{2}^{+}$ channels mainly. The reactions

\begin{eqnarray}
 H+H^{+}\rightarrow H_{2}^{+}+\gamma \\
 H+e\rightarrow H^{-}+\gamma
\end{eqnarray}
are followed by
\begin{eqnarray}
 H_{2}^{+}+H\rightarrow H_{2}+H^{+}.\\
 H^{-}+H\rightarrow H_{2}+e.
\end{eqnarray}

Due to its zero dipolar moment only quadrupolar rotational transitions are allowed, $J\rightarrow J \pm 2$, where $J$ is the quantum number for angular momentum; \citet{Abgrall}. Furthermore, due to its small moment of inertia (the smallest one between all molecules) the energy gap between its rotational quantum states, $\Delta E$, is large compared with other molecules ($\Delta E_{J\rightarrow J\pm 2}\propto 1/I$, where $I$ is the moment of inertia). The smallest energy gap is $\Delta E_{2\rightarrow 0}\approx 500 K$. With this energy difference it is very diffucult to reach temperatures below $\sim 100K$, (see \citet{Palla 1999} and references therein).

The $HD$ molecule forms through $D^{+}$ and $D$ channel mainly (see \citet{Dalgarno 1973}; \citet{Galli 2002})
\begin{eqnarray}
D^{+}+H_{2} \rightarrow H^{+}+HD,\\
D+H_{2} \rightarrow H+HD.
\end{eqnarray}
$HD$ has a moment of inertia greater than the $H_{2}$ one. Furthermore, it has a finite dipolar moment which allows internal dipolar trasitions (transitions of the kind $J\rightarrow J\pm 1$) and internal transition rates greater than in $H_{2}$, \citet{Abgrall}. Due to its small moment of inertia, the differences between its energy states are smaller than the energy differences of the $H_{2}$ molecule. All these properties make the $HD$ molecule an efficient cooler at low temperatures, below $\sim 100K$, (see \citet{NagakuraOmukai}; \citet{Ripamonti 2007}; \citet{McGreerBryan 2008}; \citet{Palla 1999} and references therein).\\
The $LiH$ molecule is formed mainly by radiative association of $Li$ and $H$ and associative detachment of $Li^{-}$ and $H$ \citep{Stancil 1996}:
\begin{eqnarray}
 Li+H & \rightarrow & LiH+\gamma.\\
 Li^{-}+H & \rightarrow & LiH+e^{-}.
\end{eqnarray}
Moreover, this molecule have both a dipolar moment and a moment of inertia larger than the ones of the $HD$ molecule. These characteristics could make the $LiH$ molecule an efficient cooler at low temperatures, but the cooling functions depend on the number density of the specie, so if the abundance of $LiH$ is too low as expected in primordial environments \citep{Stancil 1996} its cooling effect will be negligible. For a review of $Li$ chemestry see \citet{Bodo 2001} and \citet{Bodo 2003}.

In the work of \citet{Galli 1998} the effect of $HD$ and $LiH$ was included on the gas cooling. To calculate the photo-destruction ratescoefficient, for photoionization, photodetachment, and photodissociation, they assumed detailed balance with CMB photons. But, to study the effect of the first stars on the primordial gas we need the cross section for each photo-destruction process, in the spirit of \citet{Glover}. These cross sections are described below. Our current work includes the photo-destruction cross sections of both $Li$ and $Li^{-}$ and the photodisociation of $LiH$ (in its rovibrational ground state) in contrast to previous work.

We improve over recent works \citep{Glover,GA08} that have studied primordial cooling by exploring the effect of the $HD$ abundance and the inclusion of a stellar radiation field. This paper is organized as follow. In \S2 we describe both the thermal and chemical model required to follow the evolution of the gas temperature. In \S3 we present results and discussion. It includes the gas temperature evolution as a function of gas density and molecular coolers; the gas temperature evolution as a function of $HD$ abundance; the temperature evolution as a function of the ionization degree and finaly we show the effect of a star radiation field on the gas temperature. In \S4 we present the conclusions.

\section{Thermal and chemical model.}
As argued above, in a realistic cooling model of primordial gas, it is mandatory to include the molecular coolers. The main molecular coolers at low temperatures are $H_{2}$, $HD$ and $LiH$. In adition to the previous molecules, the model should include the main species created in the primordial nuclesynthesis. Our model includes 21 species:  $H$, $H^{+}$, $H^{-}$, $H_{2}$, $H_{2}^{+}$, $H_{3}^{+}$, $He$, $He^{+}$, $He^{++}$, $HeH^{+}$, $D$, $D^{+}$, $HD$, $HD^{+}$, $H_{2}D^{+}$, $Li$, $Li^{+}$, $Li^{-}$, $LiH$, $LiH^{+}$ and $e^{-}$. The reactions considered for these species are described in table \ref{tbl-4}, \ref{tbl-4bis} and \ref{tbl-5}. This table does not include the formation of $H_{2}$ by three body reaction because this reaction is not relevant at densities studied in this work.\\
The cooling processes (see table \ref{tbl-5}) considered in this work are:
\begin{itemize}
\item Collisional ionization: the gas loses energy by ionization of the different species in the environment.
\item Recombination: the gas loses energy by free electrons capture.
\item Collisional excitation: the gas loses energy when collisional excited electrons move to the unexcited states.
\item Bremsstrahlung: loss of energy due to the radiation emited by accelerated electrons.
\item Compton cooling: interchange of energy between free electrons and CMB radiation.
\end{itemize}
The molecular cooling functions are constructed as in \citet{Galli 1998}. For both $H_{2}$ and $HD$ molecules we adopted the approximate density dependent relation for the cooling functions. For LiH we adopted the low density limit cooling function. The molecular cooling functions are trated as in \citet{Puy 1993} at temperatures near $T_{CMB}$.\\
The first stars, by definition, formed in an environment without previous star. But, once population III stars are formed they can photoionize the halos where new stars will form. This process is quantified by the frequency dependent cross section, $\sigma_{A}(\nu)$, of the reaction $A^{i}+\gamma\rightarrow A^{i+1}+e^{-}$, where the specie $A^{i}$ in the $i$ ionization state moves to $i+1$ ionization state due to the interaction with photons.

The presence of a radiation field triggers a photo-destruction rate:
\begin{equation}
IR_{A}=4\pi\int_{\nu_{th}}^{\infty}\sigma_{A}(\nu)\frac{i(\nu)}{h\nu}d\nu,
\end{equation}
where $n_{A}$ is the number density of specie $A$; $i(\nu)$ is the specific intensity of radiation in the environment and $h$ is Planck's constant. The integral is calculated from the threshold frequency of ionization, $\nu_{th}$, to infinity.

The heating of the gas due to photoionizations (heat in $erg/cm^{3}s$) is given by
\begin{equation}
 \Gamma_{A}=n_{A}4\pi\int_{\nu_{th}}^{\infty}\sigma_{A}(\nu)\frac{i(\nu)}{h\nu}(h\nu-h\nu_{th})d\nu,
\end{equation}
where $h\nu_{th}$ is the threshold energy of ionization. Strictly, in the last two expressions $i(\nu)$ should be multiplied by $1-e^{-\tau}$, where $\tau=\int \sigma_{A}(\nu)n_{A}dl$ is the optycal depth. Here we assume $\tau>>1$. For example, $H$ photoionization cross section is $\sim 10^{-18}cm^{2}$, $\tau$ is greater than 1 for $dl>1-10^{-4}pc.$ If we take $n_{H}$ in the range $1-10^{4}cm^{-3}$. This distance is well below the halo scale distance.\\
The cross sections, the ionization rates and the heatings included in this work are mentioned in table \ref{tbl-6}\\
With each one of these functions we can calculate both the abundance of all species and the temperature evolution. The abundance of specie $A$ in the gas evolves following the equation
\begin{equation}
 \frac{\partial n_{A}}{\partial t}=C_{A,ij}-n_{A}D_{l,ph},
\end{equation}
where $C_{A,ij}=\Sigma_{ij}n_{i}n_{j}k_{RC,ij}$ is the creation rate of specie $A$ by $i$ and $j$ with $k_{RC,ij}$ the rate coefficient of the reaction and $D_{l,ph}$ is the destruction rate of specie $A$ by specie $l$, $D_{l}=\Sigma_{l}n_{l}k_{RC,Al}$ or by photoionization $D_{ph}=k_{IR}$. These couppled deferential equations are solved by the backward differencing formula, BDF \citep{Anninos 1997}\\
The gas temperature $T$ changes according to
\begin{equation}
 \frac{\partial T}{\partial t}=\frac{\gamma-1}{k_{B}\sum_{i}n_{i}}(\Gamma -\Lambda),
\end{equation}
where $\gamma$ is the adiabatic gas index; $k_{B}$ is Boltzmann's constant; $\Gamma$ is the photoionization heating and $\Lambda$ is the cooling by the processess mentioned before. In the range of both densities and temperatures studied here the gas can cool by molecular de-excitation. This is valid, i.e, for $H_{2}$ molecule until densities grater than $n\sim 10^{8}/x_{H_{2}} cm^{-3}$, with $x_{H_{2}}$ the $H_{2}$ fraction, at higher densities the molecule reach the LTE. 

\section{Results and discusion.}
We follow the gas temperature for different densities, different $HD$ abundances, different ionization conditions and different radiation conditions. The abundances relatives to $H$ at the begining of the evolution are (from \citet{Galli 1998} at $z\approx 10-20$): $n_{H^{+}}=10^{-4}$, $n_{H^{-}}=10^{-12}$,${H_{2}^{+}}=10^{-14}$, $n_{H_{2}}=10^{-3}$, $n_{H_{3}^{+}}=10^{-18}$, $n_{D}=4\times10^{-5}$, $n_{D^{+}}=0$, $n_{HD}=10^{-6}$, $n_{HD^{+}}=10^{-18}$, $n_{H_{2}D^{+}}=10^{-19}$, $n_{He^{+}}={He^{++}}=0$, $n_{HeH^{+}}=10^{-12}$, $n_{Li}={Li^{+}}=10^{-10}$, $n_{Li^{-}}=10^{-23}$, $n_{LiH}=10^{-21}$ and $n_{LiH^{+}}=10^{-18}$. The $H$ and $He$ densities are $\rho_{H}=0.75\rho_{tot}$ and $\rho_{He}=0.24\rho_{tot}$, where $\rho_{tot}$ is the total baryonic matter.

\begin{figure}[ht]
\includegraphics[scale=0.4,angle=0]{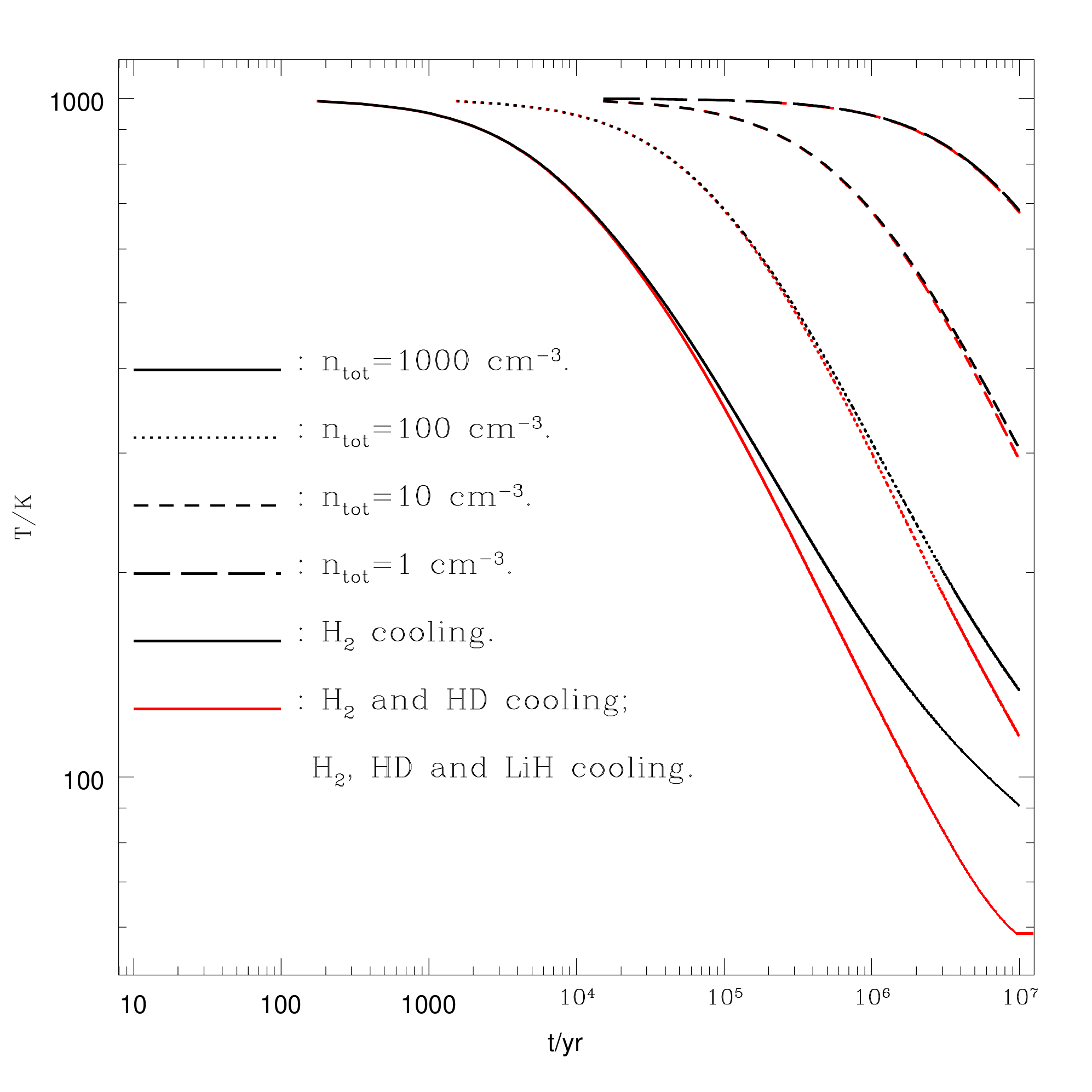}
\caption{Temperature evolution for an isolated element of gas with the initial abundances mentioned in the text. There are four diferent values of $H$ density: long dashed line $n=1 cm^{-3}$; short dashed line $n=10 cm^{-3}$; dotted line $n=100 cm^{-3}$; solid line $n=1000 cm^{-3}$. Here, different colors take into account different molecular coolers: black line are $H_{2}$ coolings; red lines are $H_{2}$ and $HD$ coolings, and $H_{2}$, $HD$ and $LiH$ coolings. The diference between the curves with $LiH$ and without $LiH$ is neglegible.\label{1}}
\end{figure}

Figure~\ref{1} shows the temperature evolution of an isolated gas element with initial temperature of $1000K$. There are four different values for $H$ abundance. These four cases show that the temperature has a strong gas density dependence; a ten times more dense gas element cools almost ten times faster in this range of density. The $LiH$ cooling seems insignificant at this temperatures and densities, whereas the effect of $HD$ cooling seems to be important at low temperatures when the elements of gas are more dense. In spite of the high $HD$ cooling efficiency, the gas element can not drop its temperature below the CMB temperature (\citet{Johnson06}; \citet{Yoshida et al07}). CMB temperature, $T_{CMB}=2.73(1+z)$, at $z=20$ is $T_{CMB}\approx 57K$. This suggests the IMF of pop III stars not depend on the gas metallicity only but on the redshift $z$ at which the stars are formed, too. For example, if we take the $T_{CMB}$ as the lowest temperature reached by the gas, $M_{J}(20)\approx 0.5\times M_{J}(30)$, with $M_{J}(z)$ the Jeans mass at redshift $z$.\\

\begin{figure}[ht!]
\includegraphics[scale=0.4,angle=0]{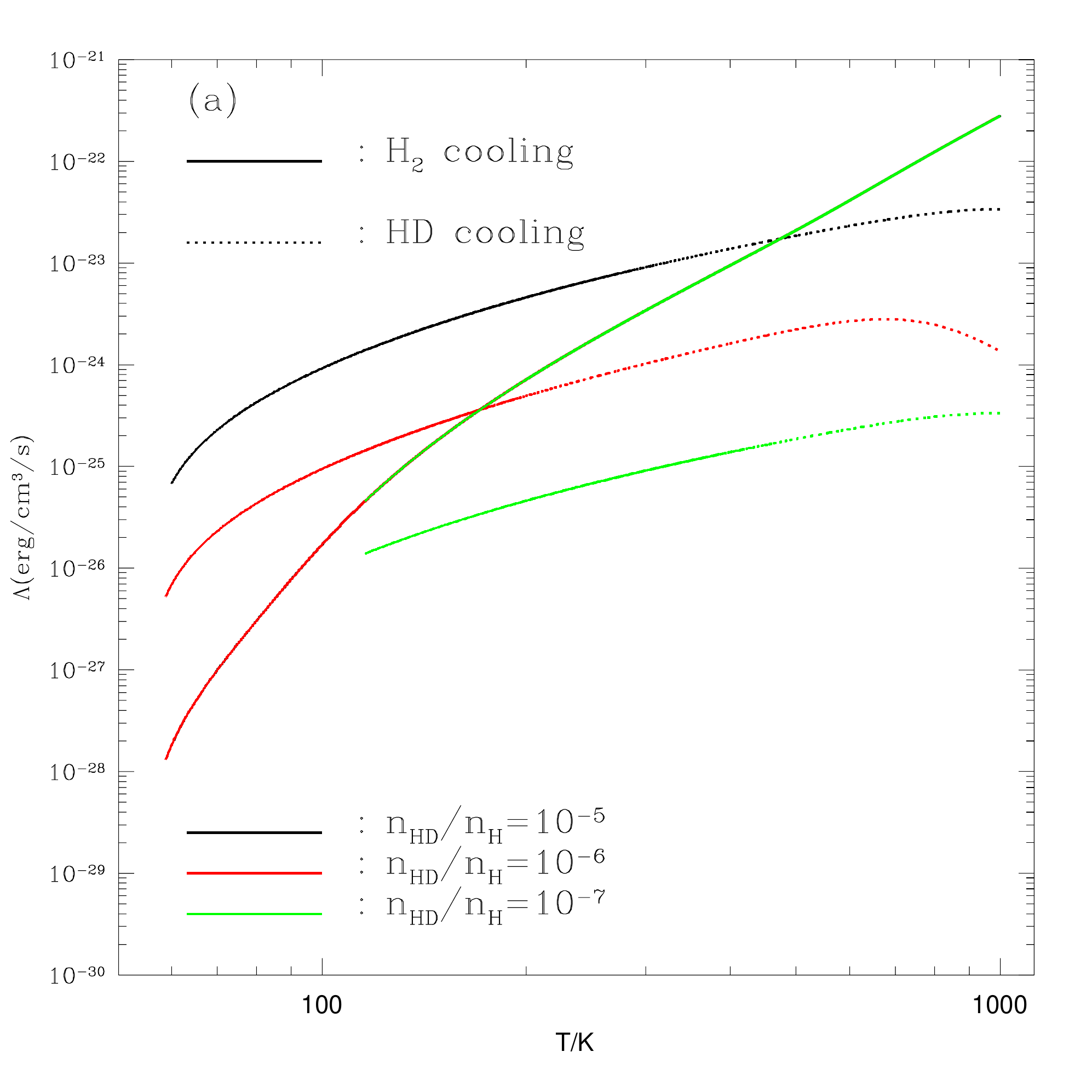}
\includegraphics[scale=0.4,angle=0]{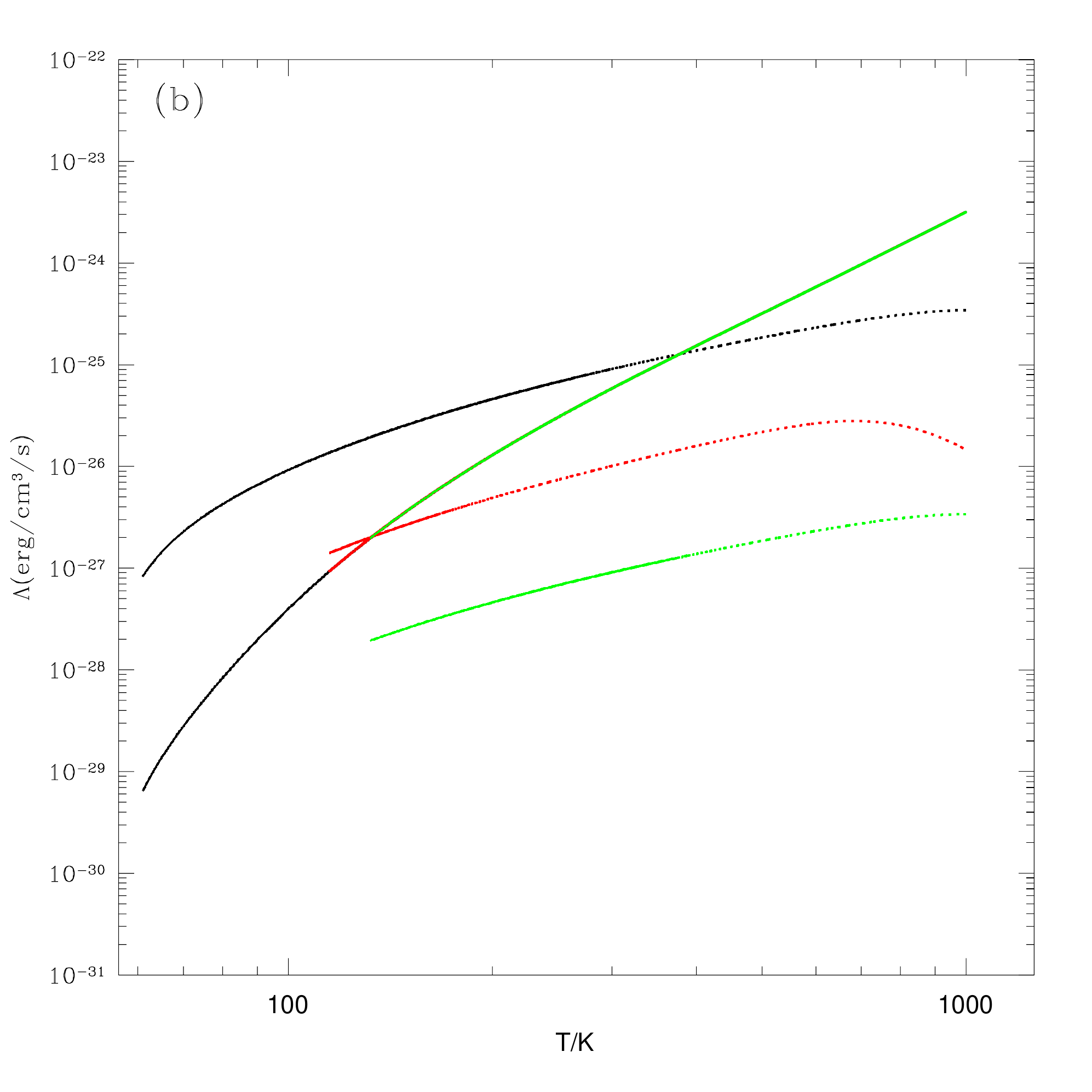}
\caption{$H_{2}$ and $HD$ at low temperatures. In panel (a) and (b) the gas number density is $n_{tot}=100-1000 cm^{-3}$, respectively. $H_{2}$ cooling in solid line; $HD$ cooling in dotted line. The diferent colors indicate diferent initial $HD$ abundances: $n_{HD}/n_{H}=10^{-5}$ in black; $n_{HD}/n_{H}=10^{-6}$ in red; $n_{HD}/n_{H}=10^{-7}$ in green.\label{2}}
\end{figure}

In order to know which cooling effect dominates the evolution of gas temperature we considered the main cooling processess independently through their thermal evolution.
Figure \ref{2} shows that the cooling of $H_{2}$, $HD$, $LiH$ -$LiH$ do not appear due to its low value- and the total cooling, $\Lambda$, divided by $n_{H}n_{H_{2}}$ for a gas element with $n_{tot}=100 cm^{-3}$ and an $H_{2}$ abundance of $n_{H_{2}}/n_{H}=10^{-3}$.
\begin{equation}
 \frac{\Lambda}{n_{H}n_{H_{2}}} \approx \lambda_{H_{2}} + \left(\frac{n_{HD}}{n_{H_{2}}}\right)\lambda_{HD} +
 \left(\frac{n_{LiH}}{n_{H_{2}}}\right)\lambda_{LiH}.
\end{equation}

The last expresion correspond to the total molecular cooling. Here $\Lambda_{A}=n_{H}n_{A}\lambda_{A}$ is the cooling due to the excitation of molecule $A$ by an $H$ atom.

This figure shows clearly that in a gas with a number density greater than $n_{tot}\approx 10^{2}cm^{-3}$ the $HD$ cooling is similar to $H_{2}$ cooling at a temperature $\sim 100K$ when $n_{HD}$ is greater than $10^{-6}n_{H}$. Then, with number densities greater than $\sim 10^{2}cm^{-3}$ the $HD$ cooling dominates the thermal evolution and the gas temperature can reach values much lower than the values reached in a gas with lower densities. This behaviour depends strongly on $HD$ abundance, which was taken in the range $10^{-5}-10^{-7}$ relative to $H$. At high densities the gas needs less $HD$ abundance to drop its temperature. \citet{Puy} found that at high densities an temperatures $\sim 200K$ the main molecular cooler is HD. This result was confirmed by \citet{Omukai}, \citet{Uehara}, \citet{Flower}, \citet{NagakuraOmukai} and \citet{Ripamonti 2007} and in a simple way with the results of figure \ref{1} and \ref{2}. On the other hand, in order to be an effecient cooler $LiH$ should increase its abundance in about ten order of magnitude due to its low abundance. This increment is too large to be real. The $LiH$ cooling is not important in these cases (see \citet{Mizusawa}).\\

\begin{figure}[ht]
\includegraphics[scale=0.4,angle=0]{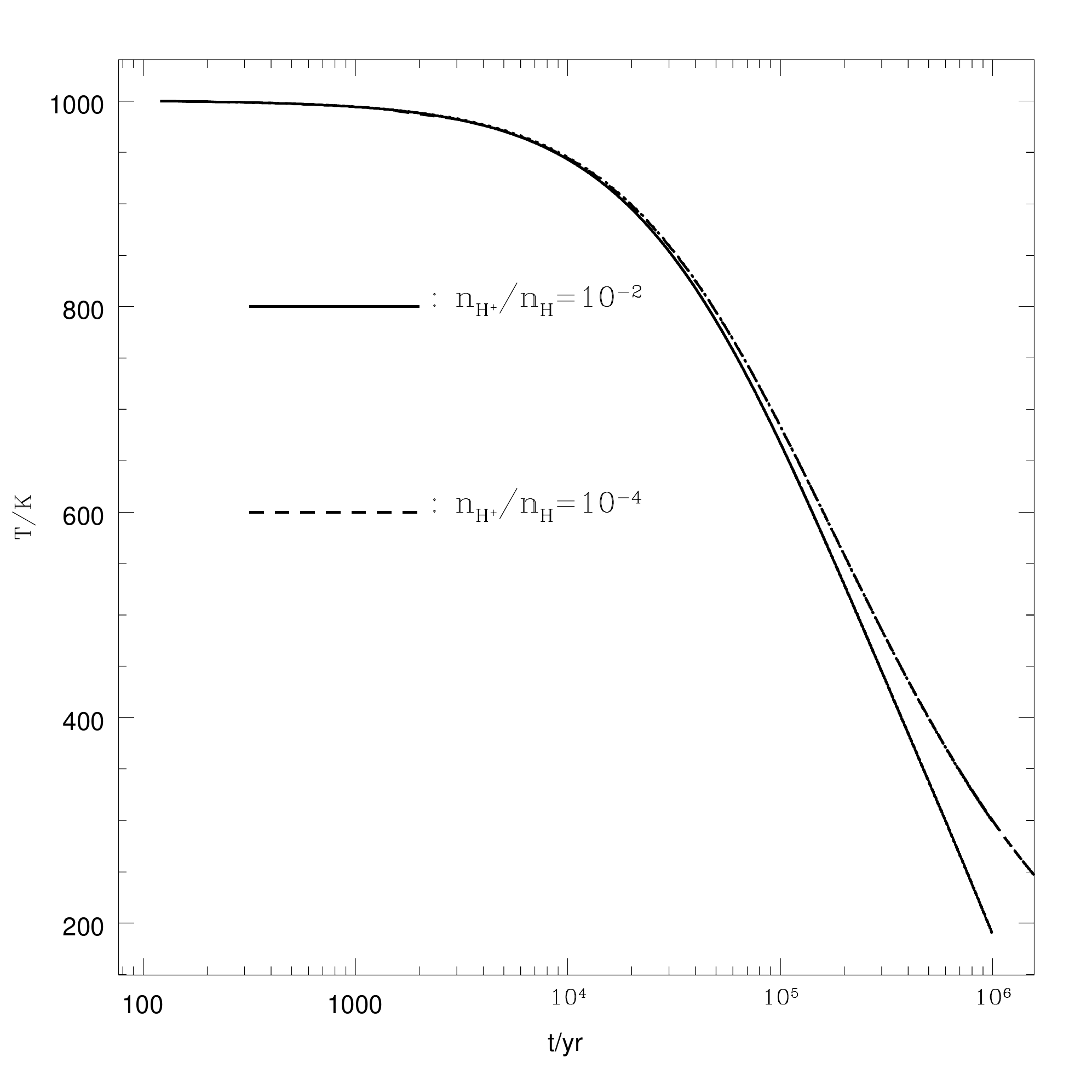}
\caption{Gas temperature as a function of the ionization degree of a gas element with $n_{tot}=100 cm^{-3}$. In solid line $n_{H^{+}}/n_{H}=10^{-2}$; long dashed $n_{H^{+}}/n_{H}=10^{-4}$.}
\label{3}
\end{figure}

Figure \ref{3} shows the gas temperature evolution of an ionized gas. This figure shows that a gas with a large ionization degree could reach lower temperatures than a gas with a small ionization degree, see for example \citet{NagakuraOmukai}. This is possible because the formation of the molecular coolers, $H_{2}$ and $HD$, need $H^{+}$, $D^{+}$ and free electrons. The temperature difference between the $n_{H^{+}}/n_{H}=10^{-4}$ and the $n_{H^{+}}/n_{H}=10^{-2}$ case is $\sim 100K$ at $100 cm^{-3}$.\\

\begin{figure}[th]
\includegraphics[scale=0.4,angle=0]{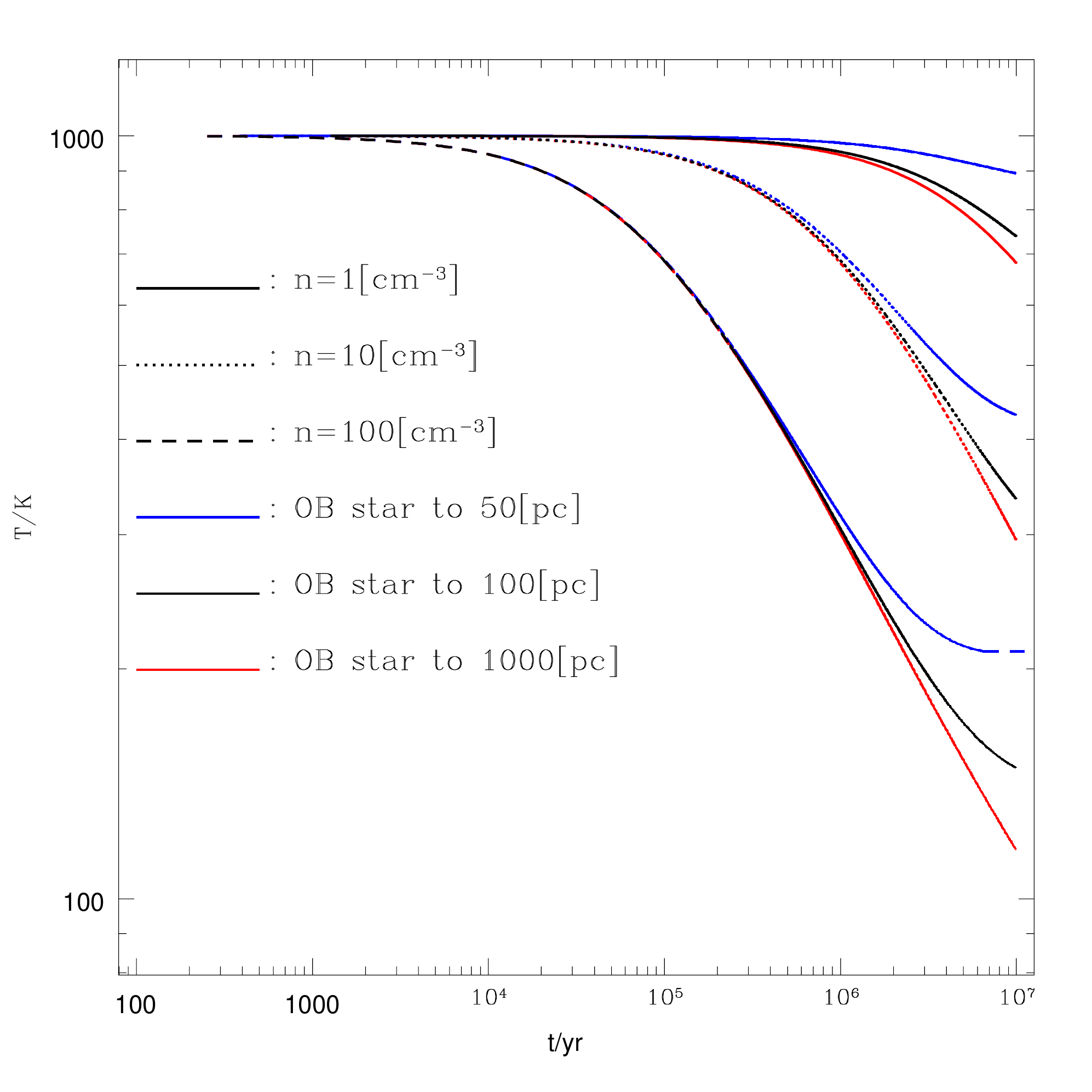}
\caption{Temperature evolution for a gas radiated by an OB star at different distances: $50pc$ in blue; $100pc$ in black; $10^{3}pc$ in red. There are three diferent values of gas density: solid line $n_{tot}=1 cm^{-3}$; dotted line $n_{tot}=10 cm^{-3}$; dashed line $n_{tot}=100 cm^{-3}$.}
\label{4}
\end{figure}

Figure \ref{4} shows the effect of an OB star ($R_{*}=15R_{\odot}$ and $T_{\rm eff}=4\times 10^{4}$) radiation field on the gas temperature at different distances. This figure shows that is very difficult for an element of gas to drop its temperature if there is an OB star at a distance less than $\sim 100pc$ due to photo-destruction of its cooling molecular agents, \citet{Yoshida et al07}. In other words if inside a halo of $\sim 10^{6}M_{\odot}$ a star is born, it is very difficult for its surounded gas reach the necesary conditions to form other stars (\citet{OhHaiman}; \citet{OmukaiNishi}). The formation of more than one star could be possible if the seed clumps of gas evolve at the same time, without radiation feedback between them. The case in which the star is at $1000pc$ and the gas has $n_{tot}=100 cm^{-3}$ present the same evolution of an isolated element of gas.

\section{Conclusions.}
In this work we have developed a model for the temperature evolution of a primordial gas including 21 different chemical species including reaction rates and cross sections available in the literature. We have paid careful attention to explore the space parameter in abundance, specially $HD$, and to include the $LiH$ to study in detail at what abundances the molecular coolants are relevant. The main results are the following:

\begin{enumerate}
\item The $HD$ molecule dominates the gas cooling at temperatures below $\sim 100-200K$ in the range of densities $10^{2}-10^{3} cm^{-3}$ for $HD$ abudances over $10^{-6}n_{H}$. The $HD$ effect is more evident at higher densities.
\item The $LiH$ molecule does not have a clear effect on the gas cooling. The gas would need an abundance at least ten orders of magnitude higher to be an efficient cooler, so $LiH$ is ruled out as an important cooler in primordial gas (\citet{Mizusawa}).
\item A gas with high ionization degree can drop its temperature more than a neutral gas because both the ionized $H$ and $D$ and electrons are catalizers in the formation of $H_{2}$ and $HD$ cooling molcules. These ionization conditions could be presents in post-shock waves zones or relic $HII$ regions, \citet{Johnson06}.
\item Is very difficult for a gas to drop its temperature in the presence of an OB star located closer than $100 pc$. So, in order to form more than one star in a primordial halo the formation of seed clumps should almost be instantaneous, otherwise the radiation feedbak of first stars will suppress the star formation conditions (\citet{OmukaiNishi}).
\end{enumerate}

This work, as previos ones,  suggests the importance of studing the effect firts stars would have on their sorrounding gas in the formation of more than one star within primordial halos (see e.g. \citet{JimenezHaiman}). For a more accurate study, we need to follow the star formation in a hydrodinamical model, which will we present in forthcoming papers.

\acknowledgments
We are grateful to CONICYT, MECESUP and FONDAP grants. The work of RJ is supported by grants from the Spanish Ministry for Science and Innovation and the European Union (FP7 program).

\clearpage

\begin{deluxetable}{llc}
\tablewidth{0pt}
\tablecaption{Rate coefficents for chemical reactions.\label{tbl-4}}
\tablehead{
\colhead{Reaction rate ($cm^{3}/s$)}           & \colhead{reaction}      &\colhead{ref}}
\startdata
Reactions with H & & \\
RC1A97   &$H+e\rightarrow H^{+}+2e$ & 39\\
RC2A97   &$H^{+}+e\rightarrow H+\gamma$ & 1\\
RC3GP98  &$H+e\rightarrow H^{-}+\gamma$ & 40\\
RC4GP98  &$H+H^{-}\rightarrow H_{2}+e$ & 18\\
RC5GP98  &$H+H^{+}\rightarrow H_{2}^{+}+\gamma$ & 10\\
RC6GP98  &$H_{2}^{+}+H\rightarrow H_{2}+H^{+}$ & 15\\
RC7GP98  &$H_{2}+H^{+}\rightarrow H_{2}^{+}+H$ & 29\\
RC8GP98  &$H_{2}+e\rightarrow 2H+e$ & 35\\
RC9A97   &$H_{2}+H\rightarrow 3H$ & 20\\
RC10A97  &$H^{-}+e\rightarrow H+2e$ & 39\\
RC11A97  &$H^{-}+H\rightarrow 2H+e$ & 1\\
RC12GP98 &$H^{-}+H^{+}\rightarrow 2H$ & 25\\
RC13GP98 &$H^{-}+H^{+}\rightarrow H_{2}^{+}+e$ & 26\\
RC14GP98 &$H_{2}^{+}+e\rightarrow 2H$ & 30\\
RC15A97  &$H_{2}^{+}+H^{-}\rightarrow H_{2}+H$ & 7\\
RC16GP98 &$H_{2}^{+}+H_{2}\rightarrow H_{3}^{+}+H$ & 38\\
RC17GP98 &$H_{2}+e\rightarrow H^{-}+H$ & 31\\
RC18GP98 &$H_{3}^{+}+H\rightarrow H_{2}^{+}+H_{2}$ & 33\\
RC19GP98 &$H_{3}^{+}+e\rightarrow H_{2}+H$ & 37\\
RC20GP98 &$H_{2}+H^{+}\rightarrow H_{3}^{+}+\gamma$ & 12\\
RC74GA08 &$H + H + H \rightarrow H_{2} + H$ & 9\\ 
RC74Inv  &$H_{2} + H \rightarrow H + H + H$ & 9\\
RC75GA08 &$H + H + H_{2} \rightarrow H_{2} + H_{2}$ & 9\\
RC75Inv  &$H_{2} + H_{2} \rightarrow H + H + H_{2}$ & 9\\
 & & \\
Reactions with He & & \\
RC21A98 & $He+e\rightarrow He^{+}+2e$ & 14\\
RC22GP98r & & 38\\
RC22A97d & $He^{+}+e\rightarrow He+\gamma$ & 4\\
RC23A97 & $He^{+}+e\rightarrow He^{++}+2e$ & 5\\
RC24GP98 & $He^{++}+e\rightarrow He^{+}+\gamma$ & 39\\
RC25GP98 & $He+H^{+}\rightarrow He^{+}+H$ & 16\\
RC26GP98 & $He^{+}+H\rightarrow He+H^{+}$ & 42\\
RC27GP98 & $He+H^{+}\rightarrow HeH^{+}+\gamma$ & 27\\
RC28GP98 & $He+H^{+}\rightarrow HeH^{+}+\gamma$ & 27, 16\\
RC29GP98 & $He+H_{2}^{+}\rightarrow HeH^{+}+H$ & 6\\
RC30GP98 & $He^{+}+H\rightarrow HeH^{+}+\gamma$ & 42\\
RC31GP98 & $HeH^{+}+H\rightarrow He+H_{2}^{+}$ & 15\\
RC32GP98 & $HeH^{+}+e\rightarrow He+H$ & 41\\
RC33GP98 & $HeH^{+}+H_{2}\rightarrow H_{3}^{+}+He$ & 23\\
\enddata
%% You can append references to a table using the \tablerefs command.
\tablerefs{(1) \citet{Abel et al. 1997}; (2) \citet{Adams 1981}; (3) \citet{Adams 1985}; (4) \citet{Aldrovandi 1973}; (5) \citet{AMDIS}; (6) \citet{Black 1978}; (7) \citet{Dalgarno 1987}; (8) \citet{Datz}; (9) \citet{FlowerHarris} ;(10) \citet{Galli 1998}; (11) \citet{Gerlich 1982}; (12) \citet{Gerlich 1992}; (13) \citet{Herbst 1982}; (14) \citet{Janev 1987}; (15) \citet{Karpas 1979}; (16) \citet{Kimura 1993}; (17) \citet{Larsson}; (18) \citet{Launay}; (19) \citet{Lepp}; (20) \citet{Mac 1986}; (21) \citet{Mielke 1994}; (22) \citet{Millar}; (23) \citet{Orient}; (24) \citet{Peart}. (25) \citet{Peterson}; (26) \citet{Poulaert}; (27) \citet{Roberge}; (28) \citet{Savin 2002}; (29) \citet{Savin 2004}; (30) \citet{Schneider}; (31) \citet{Schulz}; (32) \citet{Shavitt}; (33) \citet{Sidhu}; (34) \citet{Stancil 1993}; (35) \citet{Stibbe 1999}; (36) \citet{Stromholm}; (37) \citet{Sundstrom}; (38) \citet{Theard}; (39) \citet{Verner 1996}; (40) \citet{Wishar 1979}; (41) \citet{Yousif}; (42) \citet{Zigelman}.}
\end{deluxetable}

\begin{deluxetable}{llc}
\tablewidth{0pt}
\tablecaption{Rate coefficents for chemical reactions.\label{tbl-4bis}}
\tablehead{
\colhead{Reaction rate ($cm^{3}/s$)}           & \colhead{reaction}      &\colhead{ref}}
\startdata
Reactions with D & &\\
RC34A97 & $D+e\rightarrow D^{+}+2e$ & 14\\
RC35GP98 & $D^{+}+e\rightarrow D+\gamma$ & 1\\
RC36GP98 & $D+H^{+}\rightarrow D^{+}+H$ & 28\\
RC37GP98 & $D^{+}+H\rightarrow D+H^{+}$ & 28\\
RC38GP98 & $D+H\rightarrow HD+\gamma$ & 19\\
RC39GP98 & $D+H_{2}\rightarrow H+HD$ & 21\\
RC40GP98 & $HD^{+}+H\rightarrow H^{+}+HD$ & 15\\
RC41GP98 & $D^{+}+H_{2}\rightarrow H^{+}+HD$ & 11\\
RC42GP98 & $HD+H\rightarrow H_{2}+D$ & 32\\
RC43GP98 & $HD+H^{+}\rightarrow H_{2}+D^{+}$ & 11\\
RC44GP98 & $HD+H_{3}^{+}\rightarrow H_{2}+H_{2}D^{+}$ & 3, 22\\
RC45GP98 & $D+H^{+}\rightarrow HD^{+}+\gamma$ & 10\\
RC46GP98 & $D^{+}+H\rightarrow HD^{+}+\gamma$ &10, 34\\
RC47GP98 & $HD^{+}+e\rightarrow H+D$ & 36\\
RC48GP98 & $HD^{+}+H_{2}\rightarrow H_{2}D^{+}+H$ & 38\\
RC49GP98 & $HD^{+}+H_{2}\rightarrow H_{3}^{+}+D$ & 38\\
RC50GP98 & $D+H_{3}^{+}\rightarrow H_{2}D^{+}+H$ & 3\\
RC51GP98 & $H_{2}D^{+}+e\rightarrow 2H+D$ & 8, 17\\
RC52GP98 & $H_{2}D^{+}+e\rightarrow H_{2}+D$ & 8, 17\\
RC53GP98 & $H_{2}D^{+}+e\rightarrow HD+H$ & 8, 17\\
RC54GP98 & $H_{2}D^{+}+H_{2}\rightarrow H_{3}^{+}+HD$ & 2, 22, 13\\
RC55GP98 & $H_{2}D^{+}+H\rightarrow H_{3}^{+}+D$ & 3\\
RC76GA08 & $D + H + H \rightarrow HD + H$ & 9\\
RC76Inv  & $ HD + H \rightarrow D + H + H$ & 9\\
RC77GA08 & $ D + H + HD \rightarrow HD + HD$ & 9\\
RC77Inv  & $HD + HD \rightarrow D + H + HD$ & 9\\
\enddata
%% You can append references to a table using the \tablerefs command.
\tablerefs{(1) \citet{Abel et al. 1997}; (2) \citet{Adams 1981}; (3) \citet{Adams 1985}; (4) \citet{Aldrovandi 1973}; (5) \citet{AMDIS}; (6) \citet{Black 1978}; (7) \citet{Dalgarno 1987}; (8) \citet{Datz}; (9) \citet{FlowerHarris} ;(10) \citet{Galli 1998}; (11) \citet{Gerlich 1982}; (12) \citet{Gerlich 1992}; (13) \citet{Herbst 1982}; (14) \citet{Janev 1987}; (15) \citet{Karpas 1979}; (16) \citet{Kimura 1993}; (17) \citet{Larsson}; (18) \citet{Launay}; (19) \citet{Lepp}; (20) \citet{Mac 1986}; (21) \citet{Mielke 1994}; (22) \citet{Millar}; (23) \citet{Orient}; (24) \citet{Peart}. (25) \citet{Peterson}; (26) \citet{Poulaert}; (27) \citet{Roberge}; (28) \citet{Savin 2002}; (29) \citet{Savin 2004}; (30) \citet{Schneider}; (31) \citet{Schulz}; (32) \citet{Shavitt}; (33) \citet{Sidhu}; (34) \citet{Stancil 1993}; (35) \citet{Stibbe 1999}; (36) \citet{Stromholm}; (37) \citet{Sundstrom}; (38) \citet{Theard}; (39) \citet{Verner 1996}; (40) \citet{Wishar 1979}; (41) \citet{Yousif}; (42) \citet{Zigelman}.}
\end{deluxetable}

\begin{deluxetable}{llc}
\tablewidth{0pt}
\tablecaption{Rate coefficents for chemical reactions.\label{tbl-5}}
\tablehead{
\colhead{Reaction rate ($cm^{3}/s$)}           & \colhead{reaction}      &\colhead{ref}}
\startdata

Reactions with Li \\
RC56GP98 & $Li^{+}+e\rightarrow Li+\gamma$ & 8\\
RC57GP98 & $Li^{+}+H^{-}\rightarrow Li+H$ & 4\\
RC58GP98 & $Li^{-}+H^{+}\rightarrow Li+H$ & 4\\
RC59GP98 & $Li+e\rightarrow Li^{-}+\gamma$ & 5\\
RC60GP98 & $Li+H^{+}\rightarrow Li^{+}+H$ & 2\\
RC61GP98 & $Li+H^{+}\rightarrow Li^{+}+H+\gamma$ & 6\\
RC62GP98 & $Li+H^{-}\rightarrow LiH+e$ & 7\\
RC63GP98 & $Li^{-}+H\rightarrow LiH+e$ & 7\\
RC64GP98 & $LiH^{+}+H\rightarrow LiH+H^{+}$ & 7\\
RC65GP98 & $LiH+H^{+}\rightarrow LiH^{+}+H$ & 7\\
RC66GP98 & $LiH+H\rightarrow Li+H_{2}$ & 7\\
RC67GP98 & $Li^{+}+H\rightarrow LiH^{+}+\gamma$ & 1\\
RC68GP98 & $Li+H^{+}\rightarrow LiH^{+}+\gamma$ & 1\\
RC69GP98 & $LiH+H^{+}\rightarrow LiH^{+}+H$ & 7\\
RC70GP98 & $LiH+H^{+}\rightarrow Li^{+}+H_{2}$ & 7\\
RC71GP98 & $LiH^{+}+e\rightarrow Li+H$ & 7\\
RC72GP98 & $LiH^{+}+H\rightarrow Li+H_{2}^{+}$ & 7\\
RC73GP98 & $LiH^{+}+H\rightarrow Li^{+}+H_{2}$ & 7\\
RC78M05  & $Li+H_{2}\rightarrow LiH+H$ & 3\\
RC79M05  & $Li+H+H\rightarrow LiH+H$ & 3\\
RC80M05  & $Li+H+H_{2}\rightarrow LiH+H_{2}$ & 3\\

\enddata
%% You can append references to a table using the \tablerefs command.
\tablerefs{ (1) \citet{Dalgarno 1996}; (2) \citet{Kimura 1994}; (3) \citet{Mizusawa}; (4) \citet{Peart}; (5) \citet{Ramsbottom}; (6) \citet{Stancil Zigelman 1996}; (7) \citet{Stancil 1996}; (8) \citet{Verner 1996}.}
\end{deluxetable}

\begin{deluxetable}{llc}
\tablewidth{0pt}
\tablecaption{Cooling processess.\label{tbl-6}}
\tablehead{
\colhead{Cooling process ($erg/cm^{3}s$)}           & \colhead{reaction}      &\colhead{ref}}
\startdata
Collisional ionization & & \\
H & & \\
CI1A & $H+e\rightarrow H^{+}+2e$ & 1\\
CI2A & $H^{-}+e\rightarrow H+2e$ & 1\\
CI3A & $H^{-}+H\rightarrow 2H+e$ & 1\\
He & & \\
CI4A & $He+e\rightarrow He^{+}+e$ & 1\\
CI5A & $He^{+}+e\rightarrow He^{++}+2e$ & 1\\
D & & \\
CI6A & $D+e\rightarrow D^{+}+2e$ & 1\\
 & & \\
Recombination & & \\
H & & \\
Rec1A97 & $H^{+}+e\rightarrow H+\gamma$ & 1,2\\
He & & \\
Rec4GP98r, & & \\
Rec4A97d & $He^{+}+e\rightarrow He+\gamma$ & 1,2,3\\
Rec5GP98d & $He^{++}+e\rightarrow He^{+}+\gamma$ & 3\\
D &\\
Rec9GP98 & $D^{+}+e\rightarrow D+\gamma$ & 3\\
 & & \\
Collisional excitation of & & \\
CE1A97 & $H$ & 2,1\\
CE2GP98 & $H_{2}$ & 3\\
CE3A97 & $He$ & 2,1\\
CE4A97 & $He^{+}$ & 2,1\\
CE5GP98 & $HD$ & 3\\
CE6GP98 & $LiH$ & 3\\
 & & \\
Bremsstrahlung & & 2\\
 & & \\
Compton cooling & & 4\\
\enddata
%% You can append references to a table using the \tablerefs command.
\tablerefs{(1) \citet{Abel et al. 1997}; (2) \citet{Black 1981}; (3) \citet{Galli 1998}; (4) \citet{Peebles 1971}; (5) \citet{Verner 1996}.}
\end{deluxetable}

\begin{deluxetable}{llc}
\tablewidth{0pt}
\tablecaption{Cross sections ($\sigma_{A}$), ionization rates ($IR_{A}$) and heatings ($\Gamma_{A}$).\label{tbl-6}}
\tablehead{
\colhead{$\sigma_{A}(cm^{2})$}           & \colhead{reaction}      &\colhead{ref}\\
\colhead{$IR_{A}(s^{-1})$} & & \\
\colhead{$\Gamma_{A}(erg/cm^{-3}s)$}  &   & }

\startdata
H & & \\
$\sigma_{H}$1Gl07 & & \\
$IR_{H}$1Gl07 & & \\
$\Gamma_{H}$1Gl07 & $H+\gamma\rightarrow H^{+}+e$ & 8\\
$\sigma_{H}$2Gl07 & & \\
$IR_{H}$2Gl07 & & \\
$\Gamma_{H}$2Gl07 & $H^{-}+\gamma\rightarrow H+e$ & 2\\
$\sigma_{H}$3Gl07 & & \\
$IR_{H}$3Gl07 & & \\
$\Gamma_{H}$3Gl07 & $H_{2}+\gamma\rightarrow H_{2}^{+}+e$ & 6\\
$IR_{H}$3Gl07p & & \\
$\Gamma_{H}$3Gl07p & $H_{2}+\gamma\rightarrow H_{2}^{*}\rightarrow H+H$ & 4\\
$\sigma_{H}$Gl407 & & \\
$IR_{H}$4Gl07 & & \\
$\Gamma_{H}$4Gl07 & $H_{2}^{+} +\gamma\rightarrow H^{+}+H$ & 3\\
$\sigma_{H}$5A97 & & \\
$IR_{H}$5A97 & & \\
$\Gamma_{H}$5A97 & $H_{2}^{+}+\gamma\rightarrow 2H^{+}+e$ & 11\\
$\sigma_{H}$6A97 & & \\
$IR_{H}$6A97 & & \\
$\Gamma_{H}$6A97 & $H_{2}+\gamma\rightarrow H+H$ & 1\\
 & & \\
He & & \\
$\sigma_{He}$7Gl07 & & \\
$IR_{He}$7Gl07 & & \\
$\Gamma_{He}$7Gl07 & $He+\gamma\rightarrow He^{+}+e$ & 12\\
$\sigma_{He}$8A97 & & \\
$IR_{He}$8A97 & & \\
$\Gamma_{He}$8A97 & $He^{+}+\gamma\rightarrow He^{++}+e$ & 7\\
 & & \\
D & & \\
$\sigma_{D}$9Gl07 & & \\
$IR_{D}$9Gl07 & & \\
$\Gamma_{D}$9Gl07 & $D+\gamma\rightarrow D^{+}+e$ & 8\\
$\sigma_{D}$10A97 & & \\
$IR_{D}$10Gl07 & & \\
$\Gamma_{D}$10A97 & $HD^{+}+\gamma\rightarrow H+D^{+}$ & 11\\
$\sigma_{D}$11A97 & & \\
$IR_{D}$11Gl07 & & \\
$\Gamma_{D}$11A97 & $HD^{+}+\gamma\rightarrow H^{+}+D$ & 11\\
 & & \\
Li & & \\
$\sigma_{Li}$13RM & & \\
$IR_{Li}$13RM & & \\
$\Gamma_{Li}$13RM & $Li + \gamma\rightarrow Li^{+} + e$ & 10\\
$\sigma_{Li}$15Rams & & \\
$IR_{Li}$15Rams & & \\
$\Gamma_{Li}$15Rams & $Li^{-}+\gamma\rightarrow Li + e$ & 9\\
$\sigma_{Li}$16KD & & \\
$IR_{Li}$16KD & & \\
$\Gamma_{Li}$16KD & $LiH + \gamma\rightarrow Li + H$ & 5\\
\enddata
\tablerefs{(1) \citet{Abel et al. 1997}; (2) \citet{de Jong 1972}; (3) \citet{Dunn 1968}; (4) \citet{Galli 1998}; (5) \citet{Kirby}; (6) \citet{O'Neil}; (7) \citet{Osterbrock 1974}; (8) \citet{Osterbrock 1989}; (9) \citet{Ramsbottom}; (10) \citet{Reilman}; (11) \citet{Shapiro}; (12) \citet{Yan}.}
\end{deluxetable}

\end{document}